\documentclass{pasj00}
\begin{document}
\SetRunningHead{M.~Sato et al.}{Distance to G14.33$-$0.64}
\Received{2009/11/11}%%{yyyy/mm/dd}
\Accepted{2010/01/12}%%{yyyy/mm/dd}

\title{Distance to G14.33$-$0.64 in the Sagittarius Spiral Arm: \\ H$_2$O Maser Trigonometric Parallax with VERA}
%%% begin:list of authors
\author{
Mayumi \textsc{Sato},\altaffilmark{1,2,3}
Tomoya \textsc{Hirota},\altaffilmark{1,4}
Mark J.\ \textsc{Reid}\altaffilmark{3}
Mareki \textsc{Honma},\altaffilmark{1,4}
Hideyuki \textsc{Kobayashi},\altaffilmark{1,2,4}\\
Kenzaburo \textsc{Iwadate},\altaffilmark{1}
Takeshi \textsc{Miyaji},\altaffilmark{1}
and Katsunori M.\ \textsc{Shibata}\altaffilmark{1,4}
}

\altaffiltext{1}{Mizusawa VLBI Observatory, National Astronomical Observatory, 2-12 Hoshi-ga-oka, Mizusawa-ku, Oshu, Iwate 023-0861}
\email{mayumi.sato@nao.ac.jp}
\altaffiltext{2}{Department of Astronomy, Graduate School of Science, The University of Tokyo, 7-3-1 Hongo, Bunkyo-ku, Tokyo 113-0033}
\altaffiltext{3}{Harvard-Smithsonian Center for Astrophysics, 60 Garden Street, Cambridge, MA 02138, USA}
\altaffiltext{4}{Department of Astronomical Sciences, Graduate University for Advanced Studies, 2-21-1 Osawa, Mitaka, Tokyo 181-8588}
%%% end:list of authors
%%
%%% Please use the following style in case that sorting by 
%%% affilation is impossible. 
%%
%% \author{%
%%   D-Firstname \textsc{D-Familyname}\altaffilmark{1}
%%   E-Firstname \textsc{E-Familyname}\altaffilmark{1,2}
%%   and
%%   F-Firstname \textsc{F-Familyname}\altaffilmark{2}}
%% \altaffiltext{1}{Address of Institute}
%% \email{ddddd@xxx.xxx.xx.xx}
%% \email{eeeee@xxx.xxx.xx.xx}
%% \altaffiltext{2}{Address of Institute}
%%
%% `\KeyWords{}' always has to be placed before `\maketitle'.
%%\KeyWords{xxxx:xxxx ......} %Do NOT move this preamble from here!
\KeyWords{Galaxy: kinematics and dynamics --- Galaxy: structure --- ISM: H\emissiontype{II} regions --- ISM: individual (G14.33$-$0.64) --- masers (H$_2$O)}
\maketitle

\begin{abstract}
We report on trigonometric parallax measurements for the Galactic star forming region G14.33$-$0.64 toward the Sagittarius spiral arm.
We conducted multi-epoch phase-referencing observations of an H$_2$O maser source in G14.33$-$0.64 with the Japanese VLBI array VERA.
We successfully detected a parallax of $\pi=0.893\pm0.101$~mas, corresponding to a source distance of $d=1.12\pm 0.13$~kpc, which is less than half of the kinematic distance for G14.33$-$0.64.
Our new distance measurement demonstrates that the Sagittarius arm lies at a closer distance of $\sim1$~kpc, instead of previously assumed $\sim 2-3$~kpc from kinematic distances.
The previously suggested deviation of the Sagittarius arm toward the Galactic center from the symmetrically fitted model (Taylor \& Cordes 1993) is likely due to large errors of kinematic distances at low galactic longitudes.
G14.33$-$0.64 most likely traces the near side of the Sagittarius arm.
We attempted fitting the pitch angle of the arm with other parallax measurements along the arm, which yielded two possible pitch angles of $i=34.^\circ 7 \pm 2.^\circ 7$ and $i = 11.^\circ 2 \pm 10.^\circ 5$.
Our proper motion measurements suggest G14.33$-$0.64 has no significant peculiar motion relative to the differential rotation of the Galaxy (assumed to be in a circular orbit), indicating that the source motion is in good agreement with the Galactic rotation.

\end{abstract}

\section{Introduction} 

The Milky Way is known to be a spiral galaxy, and its structure has been intensively studied for many decades (e.g., Oort, Kerr \& Westerhout 1958; Dame \etal\ 1987, 2001).  However, there is still little agreement on the detailed spiral pattern, including the number of the spiral arms (e.g., Cohen \etal\ 1980; Drimmel 2000; Russeil 2003; Benjamin \etal\ 2005; Dame \& Thaddeus 2008; Hou \etal\ 2009).  
Spiral arms are regions of active star formation and traced primarily by H\emissiontype{II} regions, where young stellar populations (hot OB stars) ionize surrounding gas.
The major difficulty in revealing the precise spiral structure of the Galaxy arises from the lack of accurate distances to the H\emissiontype{II} regions.

Optical distance measurements such as can be obtained from photometric studies are limited in the Galactic disk by the large opacity due to dust.
Instead, the most widely used method to map the Galaxy is to adopt kinematic distances, which are derived by matching the observed radial velocities (obtained from the Doppler shift in observed frequencies) with respect to the local standard of rest (LSR) with line-of-sight velocities expected from a Galactic rotation model (e.g., Schmidt 1965; Brand \& Blitz 1993).
The famous work done by Georgelin \& Georgelin (1976) adopts this method (with the help of optical observations where available) to map H\emissiontype{II} regions in the Galaxy.
However, significant unmodelled deviations from circular motions can cause large distance errors (Burton \& Bania 1974).
Accurate and direct distance measurements without any assumption on the Galactic rotation are thus of the greatest importance to delineate the true Galactic structure.

\begin{table*}[tp]
 \begin{center}
 \caption{VERA Observations of G14.33$-$0.64}
  \begin{tabular}{clcccc}
   \hline
   \multicolumn{1}{c}{Epoch} & Date & Day of Year  &  Time Range [UT] & Beam [mas] & Beam $_{{\rm EL}>35^\circ}$ [mas]\\
   \multicolumn{1}{c}{(1)}&(2)&(3)&(4)&(5)&(6) \\
   \hline
   \hline
    1 & 2006 Oct 27  & 2006/300 & 03:00$-$12:00 & 1.87$\times$0.89 @ $-25.^\circ 5$   & 2.70$\times$0.75 @ $-7.^\circ 3$ \\
    2 & 2006 Nov 26  & 2006/330 & 01:00$-$08:45 & 1.81$\times$0.83 @ $-26.^\circ 7$   & 2.78$\times$0.73 @ $-5.^\circ 9$ \\
    3 & 2007 Jan 7   & 2007/007 & 22:00$-$05:45 & 1.87$\times$0.86 @ $-26.^\circ 1$   & --- \\
    4 & 2007 Feb 14  & 2007/045 & 20:00$-$03:43 & 1.92$\times$0.86 @ $-24.^\circ 2$   & 2.48$\times$0.81 @ $-4.^\circ 5$ \\
    5 & 2007 Mar 27  & 2007/086 & 17:00$-$00:43 & 2.10$\times$0.82 @ $-28.^\circ 3$   & 2.67$\times$0.76 @ $-5.^\circ 8$ \\
    6 & 2007 May 6   & 2007/126 & 14:00$-$21:43 & 2.24$\times$0.82 @ $-26.^\circ 0$   & 2.49$\times$0.82 @ $-8.^\circ 1$ \\
    7 & 2007 Aug 8   & 2007/220 & 08:00$-$15:50 & (1.82$\times$0.92 @ $-22.^\circ 0$) & 2.64$\times$0.77 @ $-1.^\circ 6$ \\
    8 & 2007 Oct 10  & 2007/283 & 04:00$-$11:55 & (1.72$\times$0.92 @ $-24.^\circ 6$) & 2.72$\times$0.75 @ $-3.^\circ 5$ \\
    9 & 2008 Jan 16  & 2008/016 & 21:00$-$04:55 & (1.79$\times$0.92 @ $-27.^\circ 5$) & 2.49$\times$0.81 @ $-6.^\circ 4$ \\
   10 & 2008 Apr 14  & 2008/105 & 15:00$-$22:55 & (2.00$\times$0.85 @ $-30.^\circ 9$) & 2.69$\times$0.75 @ $-8.^\circ 6$ \\
   11 & 2008 Jul 21  & 2008/203 & 08:30$-$16:15 & (1.89$\times$0.84 @ $-24.^\circ 5$) & 3.03$\times$0.70 @ $-8.^\circ 1$ \\   
   \hline
   \multicolumn{6}{@{}l@{}}{\hbox to 0pt{\parbox{150mm}{\footnotesize
(1) Epoch number.  
(2) The date of observation start time in universal time (UT).  
(3) Day of year of observation. 
(4) Start time and end time in UT.
(5) Beam size (major and minor axes) and its position angle (PA) east of north in single-beam images (with no data flagged).
    Parentheses indicate epochs not used in relative proper-motion measurements since the reference spot 4b and feature 4 were not detected. 
(6) Beam size and its PA east of north in dual-beam phase-referenced images, where data with antenna elevations below 35$^\circ$ were flagged.
     }\hss}}
   \end{tabular}
 \end{center}
\end{table*}

It has become feasible to map Galactic structure with VLBI (Very Long Baseline Interferometry) techniques, notably with the phase-referencing VLBI technique, by directly measuring trigonometric parallaxes of strong maser sources in star-forming regions associated with H\emissiontype{II} regions throughout the Galaxy.
In addition to precise distances and absolute sky positions that locate the source in 3 dimensions in the Galaxy, measurements of absolute proper motions yield the full 3-dimensional space motions (i.e., secular proper motions and source distances together give tangential velocities), which enables one to obtain full source information for Galactic structure and dynamics.
Reid \etal\ (2009b) recently refined our knowledge of the Galactic spiral structure and kinematics by integrating early results from VLBI astrometry of the Galaxy for total 18 high-mass star-forming regions (HMSFRs) with methanol, H$_2$O, SiO maser and continuum emission, carried out with the NRAO Very Long Baseline Array (VLBA) and with the Japanese VERA project.
VERA (VLBI Exploration of Radio Astrometry) is the first VLBI array dedicated to phase referencing VLBI for Galactic astrometry, consisting of 4 antennas (20~meters each in diameter) across Japan (e.g., Honma \etal\ 2000).
The recent VERA results for Galactic astrometry through maser parallax measurements are reported by Honma \etal\ (2007), Hirota \etal\ (2007, 2008a, 2008b), Imai \etal\ (2007), Sato \etal\ (2008), Kim \etal\ (2008), Choi \etal\ (2008), Nakagawa \etal\ (2008) and Oh \etal\ (2009).

The object of this study, G14.33$-$0.64 (IRAS 18159$-$1648), is a Galactic star-forming region and is VERA's first target source toward the Sagittarius spiral arm in the inner Galaxy, which is an important step toward our goal of mapping the structure of the Galaxy.

In particular, located at a low galactic longitude of $l=14.^\circ 33$ (with a latitude of $b=-0.^\circ 64$ within the Galactic plane), G14.33$-$0.64 is expected to trace the closest part of the Sagittarius arm to the Sun, and thus is an important target to determine the direct distance to the arm. 

G14.33$-$0.64 was initially discovered as a far-infrared (FIR) source in a 70-$\mu$m survey of the Galactic plane by Jaffe, Stier, \& Fazio (1982).
It was soon followed by the first detection of H$_2$O maser emission at 22~GHz associated with the FIR source by Jaffe, G\"usten, \& Downes (1981).
Later the H$_2$O maser emission was identified with an IRAS point source by Scalise, Rodriguez, \& Mendoza-Torres (1989). 
Both class I and II methanol (CH$_3$OH) maser sources were also found in the region: class II emission at 6.7~GHz (Walsh \etal\ 1995, 1997) and class I emission at 36~GHz, at 44~GHz (Slysh \etal\ 1994, 1999), at 84~GHz (Kalenski\u{\i} \etal\ 2001), and at 95 GHz (Val'tts \etal\ 2000).
G14.33$-$0.64 has been observed to display OH thermal absorption line at 1665 MHz (Wouterloot \etal\ 1993), NH$_3$ (1,1) and (2,2) inversion transition lines at 23.7~GHz (Molinari \etal\ 1996), CS($J=2\rightarrow 1$) and CS($J=5\rightarrow 4$) rotational transition lines at 98.0~GHz (Bronfman \etal\ 1996) and at 244.9~GHz (Shirley \etal\ 2003), respectively, and 1.2-mm continuum emission (Fa\'{u}ndez \etal\ 2004).
The radial velocities observed for many molecular lines of G14.33$-$0.64 are in good agreement at $V_{\rm LSR}\simeq 22$~km~s$^{-1}$.

In the present study, we report on our successful determination of the parallax of G14.33$-$0.64 with VERA as a step toward revealing the structure of the Sagittarius spiral arm in the inner Galaxy.

\section{VERA Observations}

VERA observations of the 22-GHz H$_2$O maser source (the 6$_{16}\rightarrow$5$_{23}$ rotational transition) in G14.33$-$0.64 were carried out at 11 epochs between 2006 October and 2008 July as listed in Table~1.
Using VERA's dual-beam mode for phase referencing (e.g., Honma \etal\ 2003, 2008a), we simultaneously observed the H$_2$O maser source in G14.33$-$0.64 and the extragalactic position-reference quasar (phase calibrator) J1825$-$1718 with an angular separation of 1$^\circ$.7 at a position angle (PA) of 108$^\circ$ east of north relative to G14.33$-$0.64.
The flux density of the phase calibrator J1825$-$1718 was up to $\sim$140 mJy.
A nominal H$_2$O maser position for G14.33$-$0.64 was used as reference center both for observation and for correlation: $\alpha_{2000}=$18$^{\rm h}$18$^{\rm m}$53.$^{\rm s}$8 and $\delta_{2000}=-$16$^\circ$\timeform{47'}\timeform{50.''0} (Comoretto \etal\ 1990). 
The position of J1825$-$1718 was adopted from the second VLBA Calibrator Survey by Fomalont \etal\ (2003): $\alpha_{2000}=$18$^{\rm h}$25$^{\rm m}$36.$^{\rm s}$532283 and $\delta_{2000}=-$17$^\circ$\timeform{18'}\timeform{49.''84781}.
The ICRF source NRAO~530 (J1733$-$1304; Ma \etal\ 1998) was also observed as a bright calibrator source (fringe finder) for 7-minute scans hourly in each beam.

The instrumental phase difference between the two beams were calibrated by recording the real-time phase data with an artificial noise source in each beam (Kawaguchi \etal\ 2000; Honma \etal\ 2008a). 
Left-hand circularly polarized signals were digitized at 2-bit sampling and recorded at a data rate of 1024 Mbps.
In the total bandwidth of 256~MHz (16$\times$16~MHz), one of the sixteen 16-MHz IF channels was assigned to the H$_2$O maser lines in G14.33$-$0.64.
The other 15 IF channels were for the continuum emission in the phase calibrator J1825$-$1718, with the central IF channel set at the maser frequency, using the VERA digital filter unit (Iguchi \etal\ 2005).

The data correlation was performed with the Mitaka FX correlator (Chikada \etal\ 1991).
In order to obtain sufficient resolution for the H$_2$O maser lines, only the central 8~MHz (of the 16-MHz IF channel) for the maser lines was split into 512 spectral points, yielding frequency and velocity resolutions of 15.625~kHz and 0.21~km~s$^{-1}$, respectively.
Due to the spectral splitting method, one of the other 15 IF channels for J1825$-$1718 was also split into 512 spectral points (with the maser channel), which was not used for data reduction.
The other 14 IF channels were split into 64 spectral points each and used in data reduction.

The system noise temperatures at the zenith were typically $T_{\rm {sys}}=$150$-$300~K for the first 5 epochs. 
For the last 6 epochs, one or two antennas showed higher system noise temperatures of $T_{\rm {sys}}=300-800$~K due to bad weather, while the other antennas remained at $T_{\rm {sys}}=$150$-$300~K.

\section{Data Reduction}

Visibility calibration and imaging were performed in a standard manner with the NRAO Astronomical Image Processing System (AIPS) package (Greisen 2003).
The observed frequencies of the maser lines were converted to radial (line-of-sight) velocities with respect to the local standard of rest (LSR), $V_{\rm LSR}$, using a rest frequency of 22.235080~GHz (Pickett \etal\ 1998) for the H$_2$O 6$_{16}\rightarrow$5$_{23}$ transition.

We first searched for the relative positions of all H$_2$O maser spots in the single-beam data (i.e., without phase-referencing to the calibrator J1825$-$1718 in the other beam) of the third epoch and found maser emission over several spectral components (see figure~1). 
At this epoch, the brightest H$_2$O maser channel was at $V_{\rm LSR}=26.6$~km~s$^{-1}$ (feature~7 in table~3), and the visibilities of all maser channels were firstly phase-referenced to this channel by fringe fitting (AIPS task FRING) using the channel and by applying the phase solutions to all the maser channels.
In order to find the maser spot distribution, we imaged all channels with the AIPS task IMAGR with a wide field of view of $\sim2$\timeform{''}$\times 2$\timeform{''} around the reference maser spot (feature~7), with 2048 pixels $\times$ 2048 pixels of size 1 mas.  
Many of the maser spots were outside this field ($\sim5$\timeform{''} offset from feature~7 as seen in table~3) and were found by fringe rate mapping with the AIPS task FRMAP and by shifting the image center accordingly.

\begin{figure}[t]
  \begin{center}
    \FigureFile(50mm,60mm){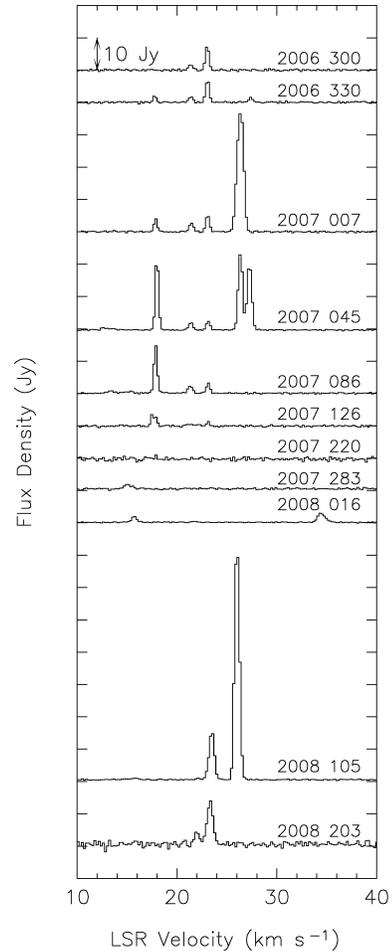}
  \end{center}
  \caption{Spectral evolution of H$_2$O maser emission in G14.33$-$0.64.  Numbers show the observed year and day of year.  Scalar-averaged cross-power spectra are shown between VERA Mizusawa and Iriki stations.  The radial velocities for many molecular lines of G14.33$-$0.64 are at $V_{\rm LSR}\simeq 22$~km~s$^{-1}$.
}\label{fig:spect}
\end{figure}

\subsection{Phase Referencing for Parallax Measurements}

%%%After finding the relative maser spot distribution at the third epoch
Next, we obtained absolute-position maps of bright maser spots by phase referencing visibilities of G14.33$-$0.64 to those of J1825$-$1718.
For each epoch, phase solutions from fringe fitting with J1825$-$1718 were applied to the H$_2$O maser channels of G14.33$-$0.64 for the corresponding frequencies.
The instrumental phase difference between the two beams was also corrected using the real-time phase-calibration data recorded during each observation.
Visibility phase errors caused by the Earth's atmosphere were calibrated based on GPS measurements of the atmospheric zenith delay which occurs due to tropospheric water vapor (Honma \etal\ 2008b). 

Since a nominal reference center of G14.33$-$0.64 was used for correlation, we first imaged the phase-referenced maser data to find the positional offset of each maser {`}feature{'}  (i.e., a group of maser spots in the same position over adjacent velocity channels) from the reference center, and then recalculated and corrected the delays for the obtained absolute positions of the maser features until the features came at the map center within 10~mas.  
After correcting the absolute position of the reference center for each maser feature, we used the same map center at all epochs for the same maser feature.
We imaged the detected maser spots with the AIPS task IMAGR for a field of view 25.6~mas $\times$ 25.6~mas around each map center, with 512 pixels $\times$ 512 pixels of size 0.05~mas.
Maser positions were fitted with elliptical Gaussian distributions with the task JMFIT.
RMS noise levels in each image per channel were typically 200$-$600 mJy/beam.

We performed least-squares fitting to simultaneously solve for the sinusoidal parallax curve and linear proper motion in right ascension (RA) for maser spots at consecutive velocity channels for two features that were persistent over more than a year.
We did not solve for the parallax in declination because positional errors due to tropospheric zenith delay residuals were larger in declination as in other measurements (e.g., Sato \etal\ 2007, 2008) and also because angular resolution was lower in declination than in RA for G14.33$-$0.64 (see table~1 for beam size), and the parallax ellipse was smaller in declination.
Instead, we removed the parallax obtained from RA fits to fit linear proper motion in declination.

Since image distortion and positional errors due to tropospheric zenith delay residuals are severe for sources at low elevation angles associated with low source declinations including G14.33$-$0.64 (e.g., Honma \etal\ 2008b), we attempted 4 different elevation cutoff values $25^\circ$, $30^\circ$, $35^\circ$, $39^\circ$, below which we flagged the data with the AIPS task UVFLG. 
For cutoffs above $39^\circ$, imaging became difficult with high sidelobes due to insufficient data.  
We adopted an elevation cutoff of $35^\circ$ to obtain the best fitting result.
For example, a typical error in the position measurement with one maser spot reduced from 0.18 mas to 0.14 mas by changing the elevation cutoff from $30^\circ$ to $35^\circ$. 
Flagging low-elevation data changed the beam size of antennas to be elongated in declination, however the beam size in RA was kept almost unchanged or slightly better (smaller) (see table~1).

\begin{table*}[t]
 \begin{center}
 \caption{Parallax fitting for G14.33$-$0.64 with elevation cutoff 35 degrees.}
  \begin{tabular}{ccclccc}
   \hline
   \multicolumn{1}{c}{Feature ID} & $V_{\rm{LSR}}$ &$N_{\rm{epochs}}$& Detected Epochs     &    RA Parallax, $\pi$  & $\mu_X$          & $\mu_Y$    \\
   \multicolumn{1}{c}{$\#$}       & [km s$^{-1}$]  &                 &                     &    [mas]               & [mas~yr$^{-1}$]  & [mas~yr$^{-1}$]  \\
   \multicolumn{1}{c}{(1)}        &      (2)       &       (3)       &   (4)               &     (5)                &  (6)             &(7) \\
   \hline
   \hline
%%%1a=ch291, 1b=ch290, 1c=ch289, 1d=ch288, 4a=ch259, 4b=ch258, 4d=ch257
1a & 14.6 & 6 & $-$$-$$-$\hspace{1.4pt}4\hspace{1.4pt}5\hspace{1.4pt}6\hspace{1.4pt}7\hspace{1.4pt}8$-$10\hspace{1.4pt}$-$ & 0.931 (0.124) &  6.13 (0.27) & $-$4.50 (0.37)\\
1b & 14.8 & 8 & $-$$-$$-$\hspace{1.4pt}4\hspace{1.4pt}5\hspace{1.4pt}6\hspace{1.4pt}7\hspace{1.4pt}8\hspace{1.4pt}9\hspace{1.4pt}10\hspace{1.4pt}11 & 0.936 (0.151) & 6.47 (0.19) & $-$4.15 (0.26)\\
1c & 15.0 & 8 & $-$$-$$-$\hspace{1.4pt}4\hspace{1.4pt}5\hspace{1.4pt}6\hspace{1.4pt}7\hspace{1.4pt}8\hspace{1.4pt}9\hspace{1.4pt}10\hspace{1.4pt}11 & 0.950 (0.141) & 6.49 (0.19) & $-$4.12 (0.26)\\
1d & 15.2 & 6 & $-$$-$$-$\hspace{1.4pt}4\hspace{1.4pt}5$-$7\hspace{1.4pt}8\hspace{1.4pt}9\hspace{1.4pt}10\hspace{1.4pt}$-$ & 1.004 (0.135) &  6.28 (0.26) & $-$4.23 (0.35)\\
1 combined &  &  & &   0.954 (0.130)     &                 &               \\
\hline
4a &  21.4 & 6 & \hspace{1.5pt}1\hspace{1.5pt}2\hspace{1.4pt}$-$\hspace{1.4pt}4\hspace{1.4pt}5\hspace{1.4pt}6$-$\hspace{0.1pt}$-$\hspace{0.1pt}$-$10\hspace{1.4pt}$-$ & 0.629 (0.171) & $-$1.60 (0.23)  & $-$0.26 (0.30)\\
4b &  21.6 & 6  & \hspace{1.5pt}1\hspace{1.5pt}2\hspace{1.4pt}$-$\hspace{1.4pt}4\hspace{1.4pt}5\hspace{1.4pt}6$-$$-$$-$10\hspace{1.4pt}$-$ & 0.631 (0.162) & $-$1.58 (0.23)  & $-$0.42 (0.30)\\
4c &  21.8 & 6  & \hspace{1.5pt}1\hspace{1.5pt}2\hspace{1.4pt}$-$\hspace{1.4pt}4\hspace{1.4pt}5$-$\hspace{0.3pt}$-$8\hspace{0.3pt}$-$10\hspace{1.4pt}$-$ & 0.900 (0.151) & $-$1.70 (0.21) & $-$0.12 (0.28)\\
4 combined     &          &         &                         &   0.768 (0.160)     &                  &               \\
\hline
1\&4 combined  &          &         &                         &   0.893 (0.101)     &                  &               \\
\hline
\multicolumn{7}{@{}l@{}}{\hbox to 0pt{\parbox{150mm}{\footnotesize
(1) Feature/spot ID, corresponding to table~3.
(2) LSR velocity of the maser spot.
(3) Total number of detected epochs.
(4) Detected epochs.
(5) Measured parallax in right ascension in mas (with estimated errors in parentheses).
(6) (7) Proper motions in right ascention and in declination, respectively.  The results presented here were obtained by fitting with a single parallax of 0.893~mas (the final result from the combined RA parallax fit).
     }\hss}}
   \end{tabular}
 \end{center}
\end{table*}

\subsection{Single-Beam Analysis for Relative Proper Motions}

We also measured relative proper motions from the single-beam data to study internal motions of H$_2$O maser spots.
Since the H$_2$O maser emission in G14.33$-$0.64 was variable over the observing period (figure~1), the phase-reference maser channel used for fringe fitting differed epoch to epoch: we used the brightest velocity channel at each epoch as the phase reference, excluding the channels around $V_{\rm LSR}\sim 26$~km~s$^{-1}$ (feature~7 in table~3) because the maser spots in this velocity range were 5\timeform{''} away from the other spots.
We imaged each maser spot with the AIPS task IMAGR for a field of view of 25.6~mas $\times$ 25.6~mas (512 pixels $\times$ 512 pixels of size 0.05 mas) by shifting the map center. 
The FWHM beam size of each epoch is shown in table 1.
RMS noise levels in each image per channel were typically 50$-$110 mJy/beam.

Maser positions were fitted with elliptical Gaussian distributions with the task JMFIT and were measured relative to the reference spot chosen at each epoch.
In order to obtain relative proper motions of all spots, we calculated all maser positions relative to the maser spot at $V_{\rm LSR}\sim 21.6$~km~s$^{-1}$ (spot~4b in table~3) by subtracting the position of this spot from the maser positions at each epoch.
Since feature~4 (including spot~4b) was only persistent over the first 6 epochs, relative proper motions were measured over the first 6 epochs. 

Our criteria for detection of a maser feature are: (1) a signal-to-noise ratio higher than 7 is obtained in the map at more than two consecutive velocity channels, (2) the spots are identified at three or more epochs for detecting relative proper motions, and (3) their positions agree with those expected from the fitted proper motions within 1 mas.  
In table~3, we also list the strong feature at $V_{\rm LSR}\sim 26$~km~s$^{-1}$, even though it has no measured proper motion.

\section{Results}

Fig~1 shows the spectral evolution of H$_2$O maser emission in G14.33$-$0.64 over the observing period: scalar-averaged spectra are shown for the baseline between VERA Mizusawa and Iriki stations.

\subsection{Parallax Measurements}

Table~2 summarizes the results from measurements of parallax $\pi$ in RA ($X$) and absolute proper motions $\mu_X$ and $\mu_Y$ in RA ($X$) and Dec ($Y$).
We used a total of seven maser spots of two maser features (features 1 and 4; feature IDs in table~2 correspond to those in table~3).
The absolute maser positions used for the measurements were: $\alpha_{2000}=$18$^{\rm h}$18$^{\rm m}$54.$^{\rm s}$67444 and $\delta_{2000}=-$16$^\circ$\timeform{47'}\timeform{50.''2640} for feature~1; $\alpha_{2000}=$18$^{\rm h}$18$^{\rm m}$54.$^{\rm s}$65341  and $\delta_{2000}=-$16$^\circ$\timeform{47'}\timeform{50.''0650} for feature~4.

Errors in the measurements are indicated in parentheses in table~2.
For single-spot measurements, errors were estimated from the residuals from the least-squares fitting with uniform weights for all epochs.
For combined fits where different spots or features were simultaneously fitted with a single parallax and with different proper motions, we have estimated the upper limit of the errors. 
We will discuss error estimates further in detail in $\S 5.1$.

The final value of the parallax (from the combined fit with all seven spots) is $\pi=0.893\pm0.101$~mas.
This corresponds to a source distance of $d=1.12\pm 0.13$~kpc.
Absolute proper motions $\mu_X$ and $\mu_Y$ of the seven spots listed in table~2 were derived using this final value of $\pi$, instead of using different $\pi$ values from single-spot measurements.

\begin{figure*}[hp]
  \begin{center}
    \FigureFile(135mm,135mm){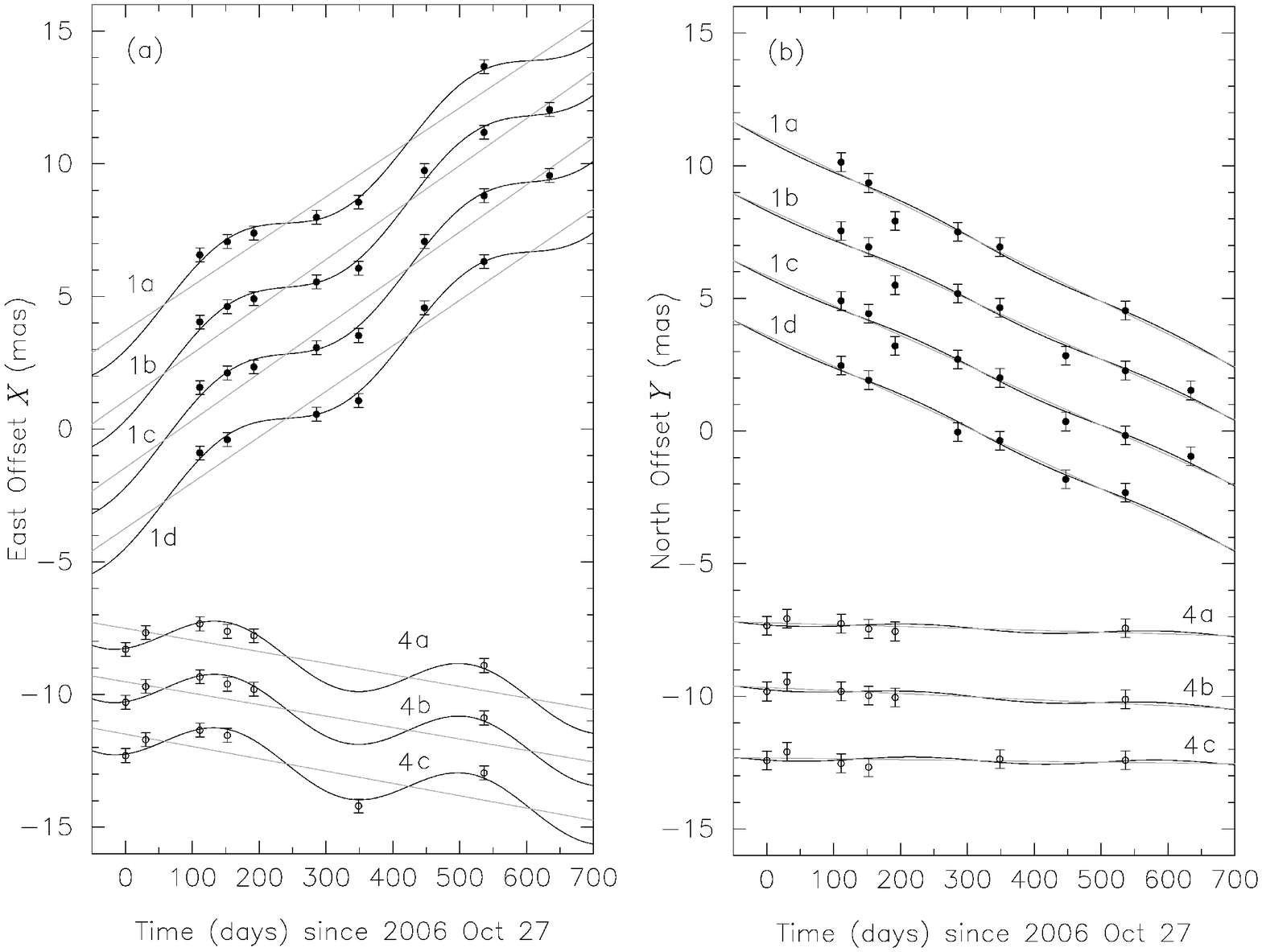}
%    %%% \FigureFile(width,height){filename}
  \end{center}
  \caption{Parallax and absolute proper-motion measurements for G14.33$-$0.64. Filled and open circles show positional evolution of maser features 1 and 4, respectively, with respect to the reference positions at origin.  (a) East offset ($X$) in mas from the reference positions of RA(J2000)$=$18$^{\rm h}$18$^{\rm m}$54.$^{\rm s}$674440  for maser feature 1 and RA(J2000)$=$18$^{\rm h}$18$^{\rm m}$54.$^{\rm s}$653410 for feature 4, as a function of time in days since the first epoch.  Best-fitting models for parallax and proper motion are shown in solid curves and gray lines, respectively. Additional shifts are given for clarity: $\Delta X=$ $+$6, 3.5, 1, $-$1.5 mas for 1a, 1b, 1c, 1d; $-$5, $-$7, $-$9 mas for 4a, 4b, 4c, respectively.
(b) North offset ($Y$) in mas from the reference position of Dec(J2000)$=-$16$^\circ$\timeform{47'}\timeform{50.''26400} for feature 1 and 
Dec(J2000)$=-$16$^\circ$\timeform{47'}\timeform{50.''06500} for feature 4, as a function of time in days since the first epoch. Additional shifts are given for clarity: $\Delta Y=$ $+$9, 6.5, 4, 1.5 mas for 1a, 1b, 1c, 1d; $-$9, $-$11.5, $-$14 mas for 4a, 4b, 4c, respectively. }\label{fig:parallax1}
\end{figure*}

\begin{figure}[th]
  \begin{center}
    \FigureFile(70mm,70mm){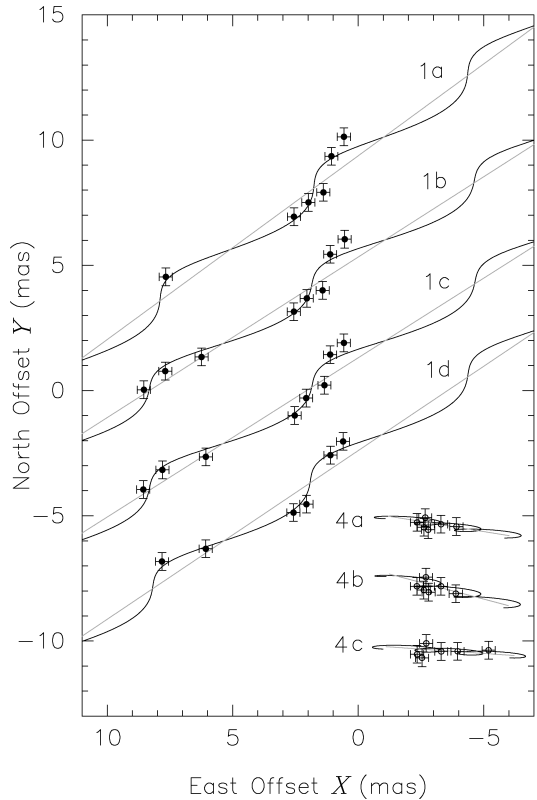}
  \end{center}
  \caption{Trajectory of maser positions on the sky.  Reference positions are the same as in figure~2 for each feature. 
Additional shifts are given for clarity: $\Delta Y=$ $+$9, 5, 1, $-$3 mas for 1a, 1b, 1c, 1d; $-$7, $-$9.5, $-$12 mas for 4a, 4b, 4c, respectively.}\label{fig:parallax2}
\end{figure}

Figures~2 and 3 show the position measurements of the seven spots used for parallax and absolute proper motion fitting.
Numbers indicate feature IDs corresponding to those in table~2.
Figures~2a and 2b show eastward ($X$) and northward ($Y$) positional offsets versus time, respectively, for seven maser spots.
Additional constant offsets are added to each maser spot in the figures for clarity.
The best-fit models for the single parallax (solid curves) and different proper motions (gray lines) are plotted for the seven spots from the combined fit.
Error bars are plotted for the standard deviation ($\sigma$) of the post-fit residuals from the least-squares fitting. 
Figure~3 shows the trajectory on the sky.

\subsection{Proper Motions}

\begin{table*}[tbh]
 \begin{center}
 \caption{Relative proper motions.}
  \begin{tabular}{ccclrrrr}
   \hline
   \multicolumn{1}{c}{Feature ID} & $V_{\rm{LSR}}$ &$N_{\rm{epochs}}$& Epochs          &    $x_1$   &  $y_1$   & $\mu_x$          & $\mu_y$    \\
   \multicolumn{1}{c}{$\#$}       & [km s$^{-1}$]  &                 &                 &    [mas]   &  [mas]   & [mas~yr$^{-1}$]  & [mas~yr$^{-1}$]  \\
   \multicolumn{1}{c}{(1)}        &      (2)       &       (3)       &   (4)           &     (5)    &   (6)    &  (7)             &(8) \\
   \hline
   \hline
                       1            &  12.5          &  3              & $--$345$-$   &     302.7  & $-$199.2 &  7.05 (0.16)     &  $-$3.69 (0.39)  \\
                       1            &  12.7          &  3              & $--$345$-$   &     302.8  & $-$199.3 &  7.08 (0.19)     &  $-$3.59 (0.11)  \\
                       1            &  12.9          &  3              & $--$345$-$   &     302.9  & $-$199.1 &  6.63 (0.10)     &  $-$3.70 (0.38)  \\
                       1            &  13.2          &  3              & $--$345$-$   &     302.8  & $-$199.0 &  7.04 (0.23)     &  $-$4.15 (0.14)  \\
                       1            &  13.4          &  3              & $--$345$-$   &     303.0  & $-$199.0 &  6.53 (0.02)     &  $-$3.86 (0.03)  \\
                       1            &  14.2          &  3              & $-$$-$$-$456 &     303.2  & $-$198.1 &  6.31 (0.41)     &  $-$6.46 (1.93)  \\
                       1            &  14.4          &  3              & $-$$-$$-$456 &     303.2  & $-$198.1 &  6.22 (0.29)     &  $-$6.65 (1.83)  \\
                       1a           &  14.6          &  4              & $--$3456     &     303.0  & $-$198.8 &  6.70 (0.53)     &  $-$5.11 (2.01)  \\
                       1b           &  14.8          &  4              & $--$3456     &     303.1  & $-$198.8 &  6.66 (0.23)     &  $-$4.22 (0.85)  \\
                       1c           &  15.0          &  3              & $--$345$-$   &     302.9  & $-$199.3 &  7.33 (0.06)     &  $-$2.14 (0.13)  \\
                       1 w-mean     &                &                 &              &     303.0  & $-$199.0 &  6.64 (0.02)     &  $-$3.79 (0.03)  \\
                       \hline
                       2            &  17.6          &  4              &  $-$23$-$56  &    187.3   & $-$134.1 & 2.23 (0.23)      &  $-$3.69 (0.19)  \\ 
                       2            &  17.8          &  5              &  123$-$56    &    187.3   & $-$134.2 & 2.12 (0.09)      &  $-$3.64 (0.10)  \\
                       2            &  18.0          &  4              &  123$--$6    &    187.4   & $-$134.2 & 2.05 (0.19)      &  $-$4.06 (0.26)  \\
                       2            &  18.2          &  3              &  123$-$$-$$-$&    187.4   & $-$134.2 & 1.41 (0.46)      &  $-$3.43 (0.28)  \\
                       2 w-mean     &                &                 &              &    187.3   & $-$134.1 & 2.10 (0.08)      &  $-$3.67 (0.08)  \\
                       \hline
                       3            &  18.0          &  4              &  $--$3456    &    183.7   & $-$129.4  &  $-$0.14 (0.39) &  $-$1.57 (0.45)   \\
                       3            &  18.2          &  4              &  $--$3456    &    183.7   & $-$129.5  &  $-$0.35 (0.23) &  $-$0.75 (0.18)   \\
                       3            &  18.4          &  3              &  $--$345$-$  &    183.8   & $-$129.6  &  $-$0.57 (0.47) &  $-$0.42 (0.69)   \\
                       3            &  18.6          &  3              &  $--$345$-$  &    183.9   & $-$129.8  &  $-$0.90 (0.28) &  { }{ }0.18 (0.80)\\
                       3 w-mean     &                &                 &              &    183.8   & $-$129.5  &  $-$0.50 (0.15) &  $-$0.80 (0.16)   \\
                       \hline
                       4            &  20.9          &  6              &  123456      &   0.1     &  $-$0.1     & $-$0.07 (0.05) & $-$0.17 (0.16)    \\
                       4            &  21.2          &  6              &  123456      &   0.0     &  $-$0.1     & $-$0.03 (0.02) & $-$0.14 (0.07)    \\
                       4a           &  21.4          &  6              &  123456      &   0.0     &  { }{ }0.0  &  0.07 (0.03)   & { }{ }0.02 (0.05) \\
                       4b           &  21.6          &  6              &  123456      &   ---     &  ---        &  ---           &  ---  \\
                       4c           &  21.8          &  5              &  12345$-$    &   0.0     &   0.0       & $-$0.01 (0.01) &  0.03 (0.07)  \\
                       4            &  22.0          &  5              &  12345$-$    &   $-$0.1  &   0.0       &   0.09 (0.07)  &  0.04 (0.07)  \\
                       4            &  22.2          &  4              &  1234$--$    &   $-$0.1  &   0.1       &   0.33 (0.19)  & $-$0.27 (0.27) \\
                       4 w-mean     &                &                 &              &   0.0     &   0.0       & $-$0.01 (0.01) & $-$0.02 (0.03) \\
                       \hline
                       5            &  21.6          &  5              &  12345$-$    &   $-$0.3  &   3.1       & 0.50 (0.16)    & 0.61 (0.25)    \\
                       5            &  21.8          &  5              &  12345$-$    &   $-$0.2  &   3.0       & 0.22 (0.03)    & 0.75 (0.08)    \\
                       5            &  22.0          &  5              &  12345$-$    &   $-$0.3  &   3.1       & 0.32 (0.10)    & 0.63 (0.08)    \\
                       5            &  22.2          &  5              &  12345$-$    &   $-$0.1  &   3.2       & $-$0.11 (0.19) & 0.61 (0.14)    \\
                       5            &  22.4          &  5              &  12345$-$    &   0.0     &   3.4       & 0.11 (0.12)    & $-$0.13 (0.64) \\
                       5 w-mean     &                &                 &              &   $-$0.2  &   3.1       & 0.23 (0.03)    & 0.67 (0.05)    \\
                       \hline
                       6            &  22.6          &  3              &  123$-$$-$$-$&   0.0     &   $-$16.0   & 0.73 (0.14)    & $-$1.33 (0.01) \\
                       6            &  22.8          &  6              &  123456      &   0.1     &   $-$16.0   & 0.10 (0.09)    & $-$1.44 (0.08) \\
                       6            &  23.1          &  6              &  123456      &   0.0     &   $-$16.0   & 0.19 (0.02)    & $-$1.26 (0.05) \\
                       6            &  23.3          &  6              &  123456      &   0.0     &   $-$15.9   & 0.25 (0.01)    & $-$1.43 (0.12) \\
                       6            &  23.5          &  6              &  123456      &   0.0     &   $-$15.9   & 0.17 (0.05)    & $-$1.24 (0.07) \\
                       6            &  23.7          &  6              &  12345$-$    &   0.0     &   $-$15.8   & 0.30 (0.12)    & $-$1.45 (0.10) \\
                       6 w-mean     &                &                 &              &   0.0     &   $-$16.0   & 0.23 (0.01)    & $-$1.32 (0.01) \\
   \hline
					   7            & 25.4$-$27.7    &                 &              &   $-$4810 &   319       &                &                \\
   \hline
                       1,23,456 u-mean &             &                 &              &           &             & 2.53           & $-$2.08        \\   
   \hline
   \multicolumn{8}{@{}l@{}}{\hbox to 0pt{\parbox{150mm}{\footnotesize
(1) Feature/spot ID.  {`}w-mean{'} notates error-weighted mean of all spots of each feature and {`}u-mean{'} refers to unweighted- mean of all features (see text).
(2) LSR velocity of the maser spot.
(3) Number of detected epochs.
(4) Detected epochs.
(5) (6) Positional offset in mas toward east ($X$) and north ($Y$), respectively, from the reference spot 4b expected from the linear fit at the first epoch. 
(7) (8) Relative proper motion of the spot in $X$ and $Y$, respectively, with respect to spot spot 4b.  Estimated errors are shown in parentheses.
     }\hss}}
   \end{tabular}
 \end{center}
\end{table*}

\begin{figure*}[tphb]
  \begin{center}
    \FigureFile(150mm,150mm){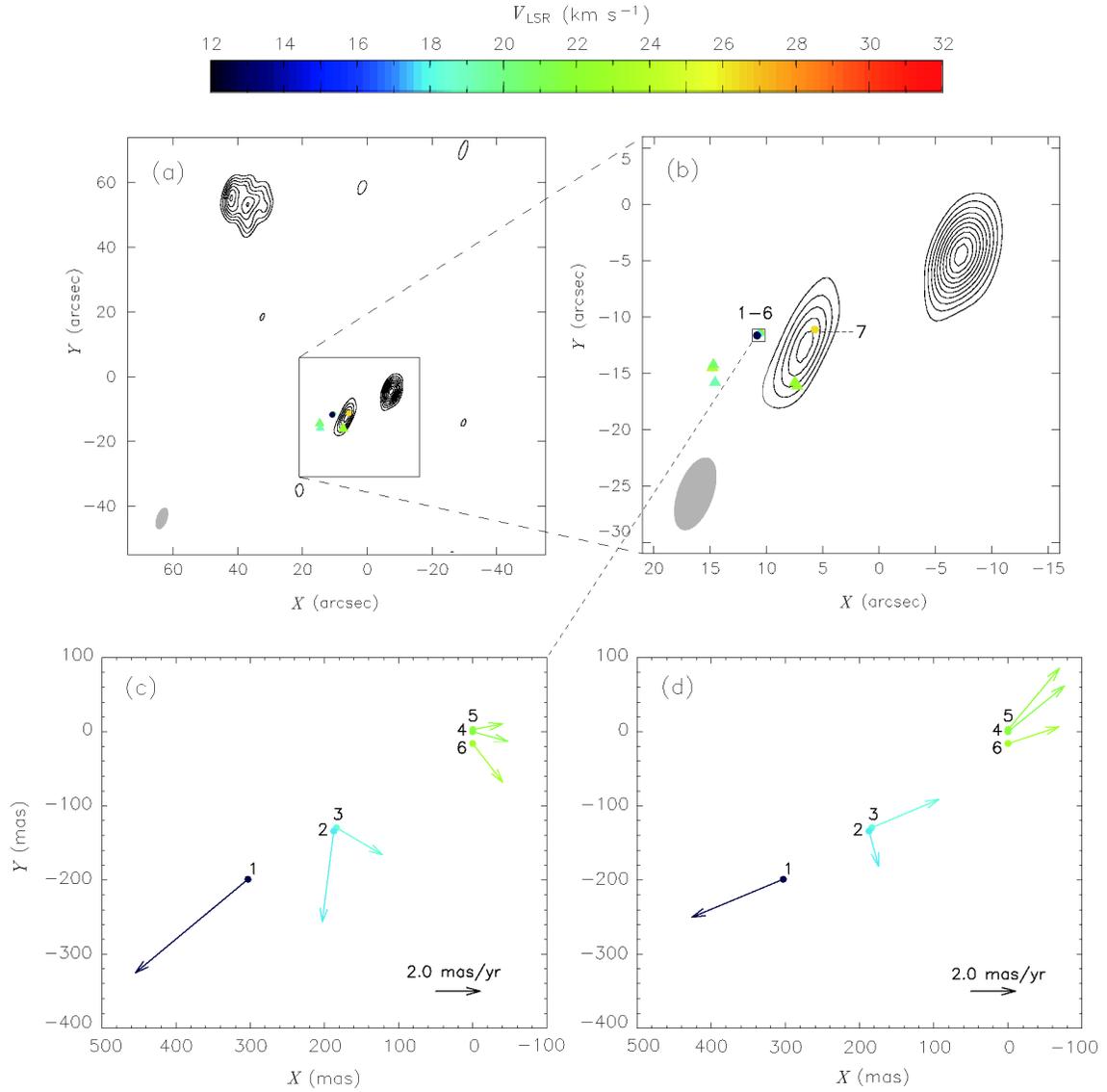}
  \end{center}
  \caption{(a) (b) Radio maps of G14.33$-$0.64 showing H$_2$O (dot) and methanol (triangle) maser positions superimposed on contours for 6-cm continuum emission obtained from VLA archive data (program AH361).
The angular resolution is \timeform{3''} for the VLA observation (Hughes \& MacLeod 1994) and the beam size is shown in gray at the left corner of each panel.
Image noise level is $\sigma=0.07$~mJy/beam, and contours are linearly spaced and correspond to 4$\sigma$, 6$\sigma$, 8$\sigma$, $\cdots$, 22$\sigma$. Peak intensity is 1.6~mJy/beam. Map origin is at the IRAS source position of $\alpha_{2000}=$18$^{\rm h}$18$^{\rm m}$53.$^{\rm s}$9, $\delta_{2000}=-$16$^\circ$\timeform{47'}\timeform{39''}. Dots represent our VLBI absolute positions of H$_2$O maser features in table~3.  Numbers correspond to feature IDs in table~3. 
Our absolute position errors essentially come from errors of the reference quasar position J1825$-$1718, which are 1.23~mas in RA and 1.97~mas in Dec (Fomalont \etal\ 2003).
Triangles show the positions of 44-GHz methanol masers mapped with the VLA by Slysh \etal\ (1999) with position errors of \timeform{0.''2}.  
Colors indicate the LSR velocity of the spots for both methanol and H$_2$O maser emission.
(c) Absolute proper motions of H$_2$O maser features without correction of the Solar motion and Galactic rotation.
Map origin (reference spot~4b) is at $\alpha_{2000}=$18$^{\rm h}$18$^{\rm m}$54.$^{\rm s}$653181 and $\delta_{2000}=-$16$^\circ$\timeform{47'}\timeform{50.''07668} (i.e., the position of spot~4b at the first epoch).
(d) Internal motions of all maser features with the mean motion of the features of ($\bar{\mu_X}$, $\bar{\mu_Y}$)$=$(0.95, $-$2.50) mas~yr$^{-1}$ removed(without correction of the Solar motion and Galactic rotation).
Map origin is the same as in figure~4c.
A proper motion of 1.00~mas~yr$^{-1}$ corresponds to a linear velocity of 5.31~km~s$^{-1}$ at a source distance of 1.12~kpc.
}\label{fig:map}
\end{figure*}

Table~3 lists the results from relative position and proper-motion measurements from the first 6 epochs.
A total of 6 maser features are presented here with proper motions detected over 3 or more epochs.

Errors of relative proper motions (shown in parentheses for $\mu_x$ and $\mu_y$ in table~3) for each spot were estimated from formal fitting uncertainties scaled by the RMS residuals of the spot positions.
For each feature ($\#$1 through 6), the relative position and proper motion were calculated as error-weighted means of those for all detected spots of the feature, as notated by {`}w-mean{'} in table~3. 

Figure~4 shows the maser distribution and proper motion for G14.33$-$0.64.
Figure~4a and 4b are radio maps of the region with contours showing continuum emission at 6-cm wavelength (C-band) from VLA archive data (program AH361) observed in the {`}C{'} configuration at an angular resolution of 3\timeform{''} (Hughes \& MacLeod 1994).
Our VLBI absolute positions of H$_2$O maser features are also shown.
Our absolute position accuracy is essentially limited by the position errors of the position-reference quasar J1825$-$1718, which are 1.23~mas in RA and 1.97~mas in Dec (Fomalont \etal\ 2003). 

Hughes \& MacLeod (1994) originally associated the H$_2$O maser emission in G14.33$-$0.64 with brighter radio continuum emission, offset \timeform{1.'16} toward northeast from the IRAS position (at the origin) as seen in figure 4a.
Our new VLBI map finds the H$_2$O maser emission (feature~7 in particular) associated (within \timeform{5''}) with a different and closer radio continuum source and yields the first precise distribution of H$_2$O maser spots in G14.33$-$0.64.

Figure~4c shows the absolute proper motions of maser features 1 to 6, which were obtained by adding relative proper motions (in table~3) to the absolute proper motion of the reference spot 4b (in table~2).
These absolute proper motions are not corrected for apparent motions due to the Solar motion and the Galactic rotation, in addition to the peculiar motion of the source.
The map origin is the position of the reference spot~4b at the first epoch: $\alpha_{2000}=$18$^{\rm h}$18$^{\rm m}$54.$^{\rm s}$653181 and $\delta_{2000}=-$16$^\circ$\timeform{47'}\timeform{50.''07668}.

Figure~4d shows the internal motions of the maser features relative to the mean motion of the features.
The mean motion of all features 1 to 6 was obtained by averaging obtained proper motions over the 3 distinct regions: (1) feature~1; (2) features~2 and 3; and (3) features~4, 5 and 6.
We took an unweighted mean of relative motions of maser features in each region, and then took an unweighted mean of the 3 regions, as listed in table~3 by {`}1,23,456 u-mean{'}.
We obtained mean relative motion of ($\bar{\mu_x}$, $\bar{\mu_y}$)$=$(2.53, $-$2.08) mas~yr$^{-1}$ (the bar symbols indicate mean values).
By adding the absolute proper motion of the reference spot 4b, ($\mu_X$, $\mu_Y$)$_{\rm{4b}}=$($-$1.58, $-$0.42) mas~yr$^{-1}$, we obtained the absolute mean motion ($\bar{\mu_X}$, $\bar{\mu_Y}$)$=$(0.95, $-$2.50) mas~yr$^{-1}$.
Note a proper motion of 1.00~mas~yr$^{-1}$ corresponds to a linear velocity of 5.31~km~s$^{-1}$ at a source distance of 1.12~kpc.
In $\S5.5$, we will adopt this mean motion to discuss the systemic motion of G14.33$-$0.64, by taking errors into account to allow for the possibility that the mean maser motion does not trace the systemic motion.

\section{Discussion}

\subsection{Astrometric Error Sources}

In this section, we will discuss possible error sources in our parallax and proper-motion measurements and how we estimated the errors.

The first possible error source in individual position measurements is thermal errors due to noise, which can be approximated by the halfwidth (HWHM) of the beam size divided by the signal-to-noise ratio of the maser map.
We find thermal errors can account for the errors of relative position measurements in the single-beam data.
Thermal errors are $\sim 0.01-0.1$~mas (i.e., beam size $\sim1$~mas, signal $\sim1-10$~Jy/beam, and noise $\sim$0.1Jy/beam) and agree well with errors in the relative proper motions as listed in table~3, which were estimated from standard deviations from the post-fit residual from the least-squares fits (see $\S3.2$ and $\S4.2$).

However, for parallax and proper-motion measurements in the dual-beam data, errors in the measurements are larger than expected from thermal errors of $\sim 0.1$~mas (i.e., beam size $\sim1$~mas, signal $\gtrsim3$~Jy/beam, and noise $\sim 0.3$~Jy/beam).
The standard deviations from the fits were $\sigma=0.26$~mas and thus are larger than thermal noise errors.

Here we do not consider the reference quasar as predominant error source since it did not show any resolved structure.
Also, even though the accuracy of the maser absolute position is limited by the uncertainties of the reference quasar position, the positional error of the reference quasar only adds as a constant offset to the maser spot position at each epoch and do not contribute to uncertainties in parallax and proper-motion measurements.  

One of the likely sources that would cause large errors in the parallax and proper-motion measurements is errors in modeling of tropospheric zenith delay (see Sato \etal\ 2008 and references therein).
Indeed, the fact that a high elevation cutoff of $35^\circ$ yielded the best-fit result for the parallax fitting for G14.33$-$0.34 indicates that this low-declination source is subject to tropospheric delay errors.
However, if errors in modeling of tropospheric zenith delay are the predominant error source, then all maser features at the same epoch should show systematic errors in the position measurements.
As can be seen in figure 2, the deviations from the parallax fits clearly differs for the two different maser features at each eposh (features 1 and 4), which indicate that errors are random for different features at the same epoch.
Therefore, it is likely that errors in modeling of tropospheric zenith delay are not the predominant error source in the remaining data after having removed as much effects of the tropospheric delay errors as possible by adopting a high elevation cutoff.

Another likely error source in the parallax measurements is variation in maser structure.
In figure 2, the tendency of deviations from the parallax fits is similar for maser spots in the same feature but different between different features (features 1 and 4).
This is consistent with the fact that the variation of maser structure causes positional errors that are uncorrelated for different features but might be correlated between maser spots in adjacent velocity channels within the same feature.
In our parallax measurements, the variation of maser structure is likely the predominant error source.

We estimated errors of the parallax measurements from the post-fit residuals from the least-squares fitting.
For different spots within the same maser feature (e.g., spots 1a, 1b, 1c, and 1d in feature \#1), we allow for the possibility that errors due to variation in maser structure data may be partially correlated.
As a conservative approach, we assumed errors of all spots within the same maser feature at the same epoch are 100\% correlated (but random for different features).
This means that, even though we used 7 maser spots of 1a, 1b, 1c, 1d, 4a, 4b, 4c for the measurements, we assume only 2 different maser features contribute as 2 independent spots to reduce the errors of the measurements.
We obtained an error of $\sigma_\pi=0.101$~mas.
Instead, if we assume errors due to variation in maser structure are random and uncorrelated for all the 7 spots (1a$-$1d and 4a$-$4c), the errors would reduce to $\sigma_\pi '=0.060$~mas.
In reality, errors of different spots in the same feature are not likely to be 100\% correlated but only partially correlated (if not uncorrelated).
Therefore, the error estimate of $\sigma_\pi=0.101$~mas in our parallax measurements is the upper limit of the errors, adopted as a conservative approach.

\subsection{Distance to the Sagittarius Spiral Arm}

\begin{figure*}[tp]
  \begin{center}
    \FigureFile(125mm,125mm){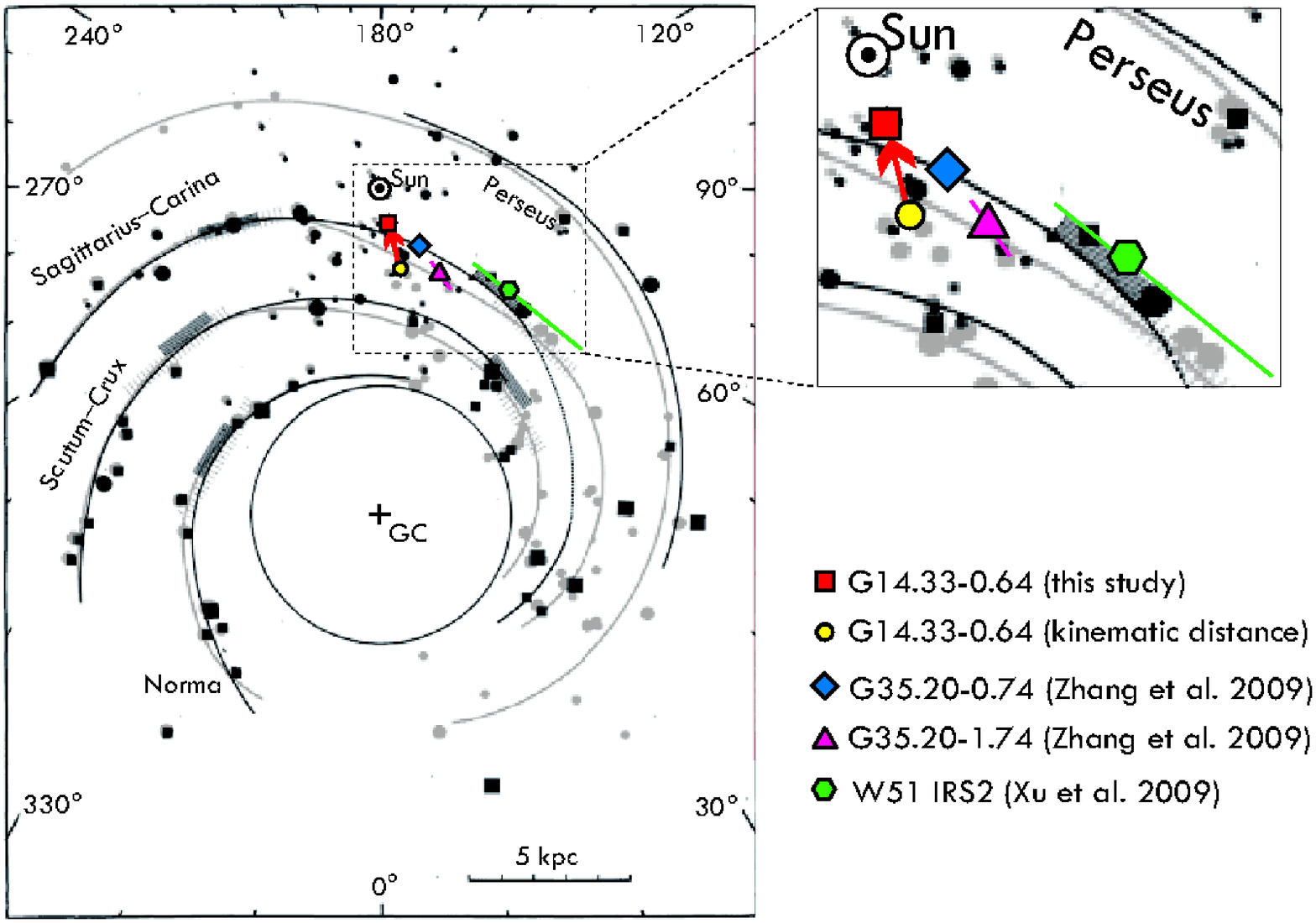}
  \end{center}
  \caption{Model of the Galaxy by Georgelin \& Georgelin (1976), overlaid with the modified model by Taylor \& Cordes (1993) shown in gray.  
   {`}$\odot${'}  indicates the location of the Sun and {`}GC{'} the position of the Galactic center.  
   The red square shows the new location of G14.33$-$0.64 based on our parallax measurements, while the yellow square is the previously estimated position of G14.33$-$0.64 based on kinematic distances. 
   Three star-forming regions, G35.20$-$0.74 (blue diamond), G35.20$-$1.74 (pink triangle) and W51~IRS2 (green hexagon), possibly belonging to the Sagittarius spiral arm are also indicated with parallactic distances measured by Zhang \etal\ (2009) and by Xu \etal\ (2009) with the VLBA for 12-GHz methanol maser emission.
   Errors for all parallactic distances are also shown, which are within the size of dots for G14.33$-$0.64 and for G35.20$-$0.74.}\label{fig:mw}
\end{figure*}
\begin{figure*}[bp]
  \begin{center}
    \FigureFile(150mm,60mm){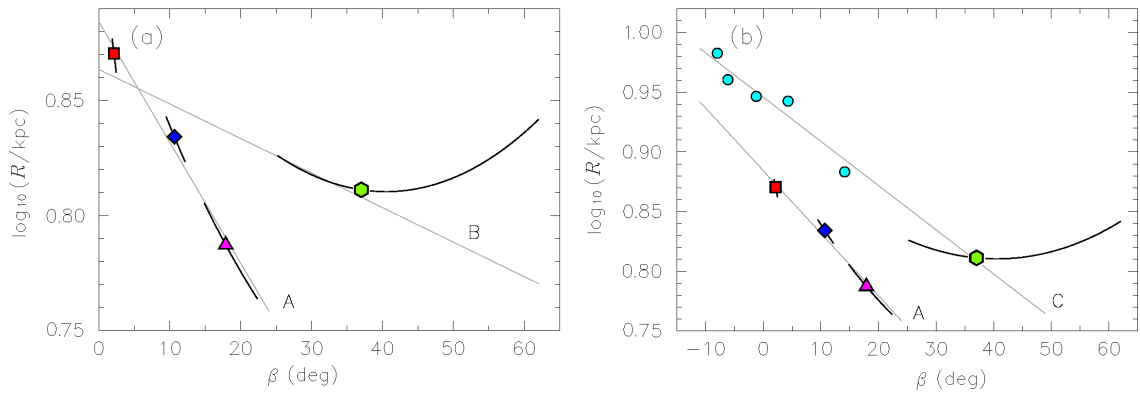}
  \end{center}
  \caption{Fits for the pitch angle of the Sagittarius spiral arm. 
The logarithm of Galactocentric radius $R$ (measured in kpc) is plotted against Galactocentric longitude $\beta$ (in degrees).
The Sun-center distance of 8.5~kpc was adopted.
(a) G14.33$-$0.64 (red square), G35.20$-$0.74 (blue diamond), G35.20$-$1.74 (pink triangle) and W51~IRS2 (green hexagon) are plotted with parallaxes and associated uncertainties from this study, Zhang \etal\ (2009) and Xu \etal\ (2009).
Gray lines show the best-fit straight lines from unweighted linear least-squares fitting to the data.
The pitch angle $i$ is obtained by taking the negative of the arctangent of the line slopes.
(Note that we need to express $\ln R$ in natural logarithm and $\beta$ in radians to calculate the pitch angle.)
Line A shows the fitting result from G14.33$-$0.64, G35.20$-$0.74 and G35.20$-$1.74, while line B is from G14.33$-$0.64, G35.20$-$0.74 and W51~IRS2.
(b) Same as (a), but with five sources (cyan dots) in the Local (Orion) arm (spur) also shown with precise parallax measurements.
Line C is an unweighted straight line fit to W51~IRS2 and the five sources in the Local arm (see text).
}\label{fig:logR}
\end{figure*}
\begin{figure}[tp]
  \begin{center}
    \FigureFile(70mm,70mm){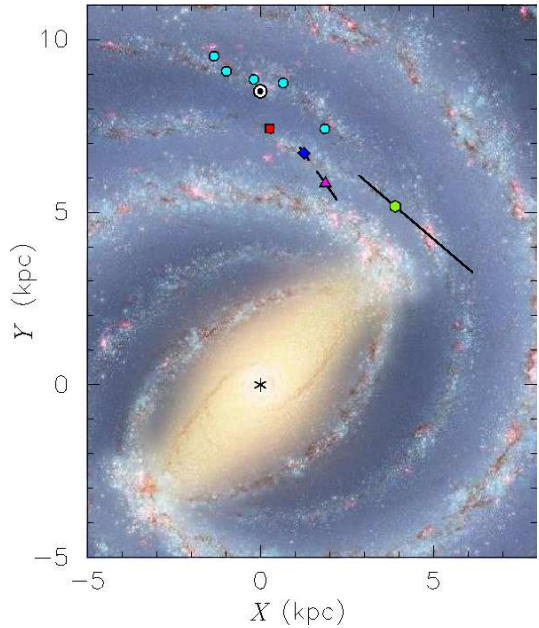}
  \caption{ 
   Galactic maser source locations in the Sagittarius and Local (Orion) arms, superimposed on artist's conception (R.\ Hurt: NASA/JPL-Caltech/SSC). 
   {`}$\odot${'} indicates the location of the Sun and {`}$*${'} the position of the Galactic center.  
   The red square shows the new location of G14.33$-$0.64 based on our parallax measurements.
   Three star-forming regions, G35.20$-$0.74 (blue diamond), G35.20$-$1.74 (pink triangle) and W51~IRS2 (green hexagon), possibly belonging to the Sagittarius spiral arm, are also indicated with parallactic distances.
   The positions of five sources in the Local (Orion) {`}arm{'} or spur are indicated by cyan dots with precise parallactic distances (see text). 
   Errors for all parallactic distances are also shown, which are mostly smaller than the size of the symbols.}\label{fig:mw_overlay}
  \end{center}
\end{figure}

Our parallax measurement for G14.33$-$0.64 reveals the source distance to be $d=1.12\pm 0.13$~kpc, which is less than half of previously derived kinematic distances.
The kinematic distances for G14.33$-$0.64 are, for example, 2.5~kpc by Molinari \etal\ (1996) from the NH$_3$ (1,1) and (2,2) lines; 2.6 kpc both by Walsh \etal\ (1997) and Val'tts \etal\ (2000) from 6.7-GHz and 95-GHz methanol maser lines, respectively; 3.1~kpc from the H110$\alpha$ line and 2.6~kpc from H$_2$CO absorption lines by Sewilo \etal\ (2004).
All of the kinematic distances above were derived using the Galactic rotation model by Brand \& Blitz (1993).
Palagi \etal\ (1993) derived 2.7~kpc from H$_2$O maser lines with a peak at $V_{\rm LSR}=22.8$~km~s$^{-1}$ using the rotation curve of Brand (1986).
The good agreement among previous kinematic distances is a result of using the same rotation model and similar radial velocities $V_{\rm LSR}\simeq 22$~km~s$^{-1}$ observed at different wavelengths.
The most persistent H$_2$O maser feature in our measurements, feature~4, also showed a radial velocity of $V_{\rm LSR}\simeq 22$~km~s$^{-1}$, which agrees well with the systemic radial velocity of G14.33$-$0.64, but several other spectral components differed up to 10~km~s$^{-1}$ in radial velocities.

Figure~5 shows the classic model of the Galaxy by Georgelin \& Georgelin (1976).
Gray lines show the modified model by Taylor \& Cordes (1993).
Note that a shift toward the Galactic center in the position of the Sagittarius arm was introduced by Taylor \& Cordes (1993), to correspond better with the kinematic distances of Downes \etal\ (1980).
Downes \etal\ (1980) estimated the kinematic distances to Galactic H\emissiontype{II} regions from radio observations of H110$\alpha$ and H$_2$CO lines using the Schmidt (1965) model, with typical errors of $\pm1$ to 2 kpc for galactic longitudes $l=20^\circ$ to $60^\circ$, which can be more than $\sim$50\% errors for the Sagittarius arm at lower galactic latitudes.
Although G14.33$-$0.64 was not in the catalog by Downes \etal\ (1980), it was in the catalog by Sewilo \etal\ (2004) in H110$\alpha$ and H$_2$CO line observations.
It can be clearly seen in figure~5 that the kinematic distance (shown as the yellow square) places G14.33$-$0.64 as well as the interpolated Sagittarius arm further toward the Galactic center, like other sources in the arm.

However, our direct parallax measurements (red square in figure~5) reveals the location of G14.33$-$0.64 to be closer to the Sun and outward the Galaxy, in good agreement with the Sagittarius arm originally modeled by Georgelin \& Georgelin (1976), without the {`}bump{'} toward the Galactic center.
In figure~5, three other star-forming regions, G35.20$-$0.74 (blue diamond), G35.20$-$1.74 (pink triangle) and W51~IRS2 (green hexagon), possibly in the Sagittarius spiral arm, are also plotted with parallax distances of $2.19^{+0.24}_{-0.20}$~kpc, $3.27^{+0.56}_{-0.42}$~kpc (Zhang \etal\ 2009) and $5.1^{+2.9}_{-1.4}$~kpc (Xu \etal\ 2009) from the VLBA for 12-GHz methanol maser emission.

It is most likely that the {`}bump{'} in the Sagittarius spiral arm toward the Galactic center suggested in Taylor \& Cordes (1993) is due to errors of kinematic distances.
A more recent model by Cordes \& Lazio (2002), which is built upon the Taylor \& Cordes (1993) model, also retains the {`}bump{'} of the Sagittarius arm toward the Galactic center.
Both Taylor \& Cordes (1993) and Cordes \& Lazio (2002) give models for the distribution of free electrons in the Galaxy, upon which most pulsar distances are determined using the observed dispersion measures (DM), i.e., the column density of electrons toward the pulsars (Frail \& Weisberg 1990).
These models are built by numerically fitting predicted and observed dispersion measures for pulsars with known {`}independent distance estimates{'} (Taylor \& Cordes 1993), most of which come from uncertain kinematic distances.

In particular, kinematic distances are more severely affected by errors of the radial velocities for sources at low galactic longitudes than at high longitudes.
For example, for the simplest assumption of circular Galactic rotation with a source distance $d$ in the solar neighborhood ($d\ll R_0$, where $R_0$ is the distance to the Galactic center from the Sun), the kinematic distance $d$ can be approximated by $d_{\rm kin}\approx V_{\rm LSR}/(A\sin(2l))$ using Oort's constant $A$ (see e.g., Karttunen \etal\ 2007).
Errors in the kinematic distances $\sigma_{d_{\rm kin}}$ are thus proportional to the errors in the radial velocities divided by $\sin (2l)$: $\sigma_{d_{\rm kin}} \propto \sigma_{V_{\rm LSR}}/\sin(2l)$.
Therefore the kinematic distances toward the Sagittarius arm in the inner Galaxy are expected to be particularly uncertain.

Taylor \& Cordes (1993) acknowledge that pulsar distances derived from previous models generally tend to be overestimated for $|l|<30^\circ$ and underestimated for $l=50^\circ - 70^\circ$ (although they claim their own model has no significant dependence of distance errors on $l$), which can account for the {`}bump{'} of the Sagittarius arm toward the Galactic center at low galactic longitudes. 
Our results as shown in figure~6 indicate that the previously expected {`}bump{'} in the Sagittarius arm toward the Galactic center is most likely due to errors that arise from kinematic distances.

Russeil (2003) points out that the nearest part of the Sagittarius arm is placed at $\sim 2$~kpc based on kinematic distances (using the rotation curve of Brand \& Blitz 1993), while a fitted regular logarithmic arm, also based on kinematic distances, passes at $\sim 1$~kpc, indicating the possibility that the Galaxy does not have a regular design.
However, our parallax measurements suggest that the nearest part of the Sagittarius arm, indeed, lies at $\sim 1$~kpc.
The disagreement between the arm fitting and the kinematic distance is likely due to errors of kinematic distances, rather than an irregular design of the Sagittarius arm.

Direct determination of distances are of great importance and required to obtain a true map of the Galaxy and, in particular, of the Sagittarius arm.
Our parallax measurement of G14.33$-$0.64 with VERA reveals the location of the Sagittarius arm to be closer to the Sun than previously thought.

\subsection{Pitch angle of the Sagittarius arm}

We attempted to fit the pitch angle $i$ of the Sagittarius arm using our parallax measurement of G14.33$-$0.64 with three other parallax measurements of sources shown in figure~5, which may lie in the Sagittarius arm: G35.20$-$0.74, G35.20$-$1.74 (Zhang \etal\ 2009) and W51~IRS2 (Xu \etal\ 2009).
The pitch angle $i$ is defined as the angle between the arm and the tangent to a Galactocentric circular orbit.
For an ideal logarithmic spiral arm, it can be expressed as, $\ln (R_1/R_2) = - (\beta_1 - \beta_2) \tan i$, for two sources 1 and 2 (indicated by subscripts) in the arm, where $R$ is the Galactocentric radius at a Galactocentric longitude $\beta$ (0 toward the Sun and increasing with galactic longitude; see Reid \etal\ 2009b).

Figure~6a shows a plot of $\log_{10} (R/{\rm kpc})$ vs.\ $\beta$ (in degrees) for G14.33$-$0.64 (red square), G35.20$-$0.74 (blue diamond), G35.20$-$1.74 (pink triangle) and W51~IRS2 (green hexagon).
Here we adopted the Sun-center distance of $R_0 =8.5$~kpc.
Errors are indicated for each source with parallax uncertainties of $\pm 1\sigma$ from this study, Zhang \etal\ (2009) and Xu \etal\ (2009) .
We attempted linear least-squares fitting to the sources with unweighted straight lines.
(Note that we need to express $\ln R$ in natural logarithm and $\beta$ in radians to calculate the pitch angle.)

As seen in figure~6a, the four sources do not lie in a straight line, and we attempted fitting with two possible combinations of three sources, which are shown in gray lines in the figure.
Line A shows a best-fit straight line for G14.33$-$0.64, G35.20$-$0.74 and G35.20$-$1.74 (excluding W51~IRS2), which yields a pitch angle of $i=34.^\circ 7 \pm 2.^\circ 7$.
Line B is a fitting result from G14.33$-$0.64, G35.20$-$0.74 and W51~IRS2 (excluding G35.20$-$1.74), which yields a smaller pitch angle of $i = 11.^\circ 2 \pm 10.^\circ 5$.
This pitch angle $i\sim 11^\circ$ agrees well with the four-arm Milky Way model by Vall\'ee (1995) with a best-fit pitch angle of $i=12.^\circ 1\pm1$.  

Figure~7 is a plot of the positions of the four sources superimposed on artist's conception of the Milky Way.
For comparison, five sources in the Local (Orion) {`}arm{'} or spur are also shown with precise parallax measurements: G59.7$+$0.1 (Xu \etal\ 2009), Cep~A (Moscadelli \etal\ 2009), Orion (Hirota \etal\ 2007; Menten \etal\ 2007; Kim \etal\ 2008), G232.6+1.0 (Reid \etal\ 2009a), and VY~CMa (Choi \etal\ 2009; Reid \etal\ 2009c). 
With the five sources, Reid \etal\ (2009b) fitted the pitch angle of the Local arm to be $27.^\circ 8 \pm 4.^\circ 7$, which is larger than the pitch angles they fitted for another spiral arm, e.g., $16.^\circ 5 \pm 3.^\circ 1$ for the Perseus spiral arm.

In figure~6b, we also attempted a straight line fitting (line C) to the five Local arm sources (marked by cyan dots) plus W51~IRS2 (green hexagon), which yields a pitch angle of $26.^\circ 1 \pm 12.^\circ 3$, which is consistent with the pitch angle fitted with only five sources above.
Thus the Local arm/spur may branch from the Sagittarius arm near the position of W51~IRS2, which is often considered to be at the tangent point of the Sagittarius arm.
One possible interpretation is that the Sagittarius arm bifurcates near the position of W51~IRS2 into the Local spur (line C) at a pitch angle of $i\sim 26^\circ$ and into the other arm traced by G14.33$-$0.64 and G35.20$-$0.74 (line B) at a pitch angle of $i\sim11^\circ$.
Another possibility is that the Sagittarius arm is traced by G14.33$-$0.64, G35.20$-$0.74 and G35.20$-$1.74 (line B) and branches from the interior (Scutum-Crux) arm at a large pitch angle of $i\sim34^\circ$.
However, more sources with precise parallaxes are needed to establish clear spiral arm structure.
Ongoing and future parallax measurements with VERA and with the VLBA are expected to reveal the structure of the Sagittarius arm and other spiral arms of the Galaxy further in detail.

\subsection{Magnetic Field Reversals and the Sagittarius Arm}

It is of interest to compare our results for the distance to the Sagittarius arm with studies of Galactic magnetic field reversals.
The Galactic magnetic field has been probed most often by Faraday rotation measure (RM) observations of linearly polarized emission from both pulsars (e.g., Noutsos \etal\ 2008) and extragalactic radio sources (e.g., Brown \etal\ 2007).
A common conclusion in many pulsar polarization studies is that the magnetic field in the Local arm is clockwise while it is counterclockwise in the first quadrant ($0^\circ \le l \le 90^\circ$) component of the Sagittarius arm, indicating the existence of a magnetic field reversal between the arms.  
Weisberg \etal\ (2003) found from pulsar polarimetry a null in the magnetic field of a width less than 0.5~kpc extending from near the Sun over 7~kpc toward $l\sim 60^\circ$ (figure~4 in Weisberg \etal\ 2003), located midway between the Local and Sagittarius arms, which is most likely the field reversal region.

Weisberg \etal\ (2003) noted a "1-kpc wide strip" of steady magnetic field from the local reversal (midway between the Local and Sagittarius arms) into the Sagittarius arm, based on the Sagittarius arm model by Cordes \& Lazio (2002).
As previously discussed, our parallax measurements demonstrate the Sagittarius arm lies at a closer distance of $\sim1$~kpc, instead of previously estimated $\sim 2-3$~kpc from kinematic distances, and we find that G14.33$-$0.64 (this study) and G35.20$-$0.74 (Zhang \etal\ 2009) trace out the near side of the Sagittarius arm, which lie outside of the {`}bump{'} delineated in Taylor \& Cordes (1993) as well as in Cordes \& Lazio (2002).
Our parallax measurements thus indicate that the strip of steady magnetic field found by Weisberg \etal\ (2003) is likely in the Sagittarius arm, rather than in an inter-arm region exterior to the arm.
This lends support to the fact that the magnetic field in the Sagittarius arm is steadily and dominantly counterclockwise, and is further evidence for the conclusion of Weisberg \etal\ (2003) that the field maxima tend to lie along the spiral arms, while the field reversals occur between the arms.

\subsection{Motion of G14.33$-$0.64 and the Galactic Rotation}

As seen in figures 4c and 4d, the internal motions of the H$_2$O masers in G14.33$-$0.64 show a bipolar jet-like motion on the sky, with deviations of $\simeq 1-2$~mas~yr$^{-1}$ from the mean, which correspond to a linear velocity of $5-10$~km~s$^{-1}$ at a distance of 1.12~kpc.
The central radial velocity $V_{\rm LSR}\simeq 22$~km~s$^{-1}$ of the maser emission agrees well with other molecular line velocities, and the deviations up to 10~km~s$^{-1}$ from the central radial velocity agree with the proper motions.

From the parallax, proper motion, radial velocity and the sky position of the H$_2$O maser source, we can now calculate the full three-dimensional position and velocity of the source in the Galaxy.
By following the methods described in detail by Reid \etal\ (2009b) to convert from the heliocentric reference frame to a reference frame that rotates with the Galaxy, we obtain the peculiar motion of the source with respect to the Galactic rotation.

We adopt the mean absolute proper motion (the reference frame in figure~4d) of ($\bar{\mu_X}$, $\bar{\mu_Y}$)$=$(0.95, $-$2.50) mas~yr$^{-1}$ as the systemic motion of the source (before the correction of the solar motion and the Galactic rotation), with uncertainties of $\pm$2~mas~yr$^{-1}\simeq \pm$10~km~s$^{-1}$ in each of the eastward ($X$) and northward ($Y$) directions to allow for the possibility that the mean maser motion does not trace the systemic motion.

For the radial velocity, we adopted $V_{\rm LSR}=22\pm 10$~km~s$^{-1}$.
Adopting the {\it Hipparcos} solar motion values of $U_\odot=10.0\pm 0.36$~km~s$^{-1}$ (radially toward the Galactic center), $V_\odot=5.25\pm0.62$ (in the local direction of Galactic rotation) and $W_\odot = 7.17\pm0.38$~km~s$^{-1}$ (vertically upwards, i.e., toward the north Galactic pole perpendicularly to the Galactic plane) from Dehnen \& Binney (1998) with the recent best-fit results for the Galactic constants of $R_0 = 8.4\pm0.6$~kpc and $\Theta_0 = 254\pm16$~km~s$^{-1}$ by Reid \etal\ (2009b), and assuming a flat rotation of the Galaxy (i.e., rotational velocity $\Theta$ at the source location is the same as at the Sun, $\Theta \simeq \Theta_0$) the peculiar velocity components of G14.33$-$0.63 are obtained to be $U_s = 11\pm10$~km~s$^{-1}$ toward the Galactic center at the source position, $V_s = -1 \pm 11$~km~s$^{-1}$ in the local direction of the Galactic rotation, and $W_s = -4 \pm 11$~km~s$^{-1}$ vertically out of the Galactic plane toward the north Galactic pole.

Here the uncertainties of $10-11$~km~s$^{-1}$ in the derived peculiar motion are directly due to the uncertainties for the proper motion and radial velocity of G14.33$-$0.64.
The contribution from uncertainties in the Galactic constants $R_0$ and $\Theta_0$ are negligible, because the Galactic rotation term is almost canceled out in the differential calculation. 
If we adopt the IAU standard values of $R_0 = 8.5$~kpc and $\Theta_0 = 220$~km~s$^{-1}$ instead, the resulting peculiar motion becomes  $U_s = 12\pm10$~km~s$^{-1}$, $V_s = -1 \pm 11$~km~s$^{-1}$, and $W_s = -4 \pm 11$~km~s$^{-1}$.
Therefore, the peculiar motion of G14.33$-$0.64 is not significant in the direction of Galactic rotation ($V_s$) or in the direction out of the Galactic plane ($W_s$).
For the source location of G14.33$-$0.64 relative to the Sun in the Galaxy, the larger peculiar velocity component of G14.33$-$0.64 toward the Galactic center ($U_s$) reflects a radial velocity larger than expected from the circular rotation model, which has led to the larger kinematic distances derived in the previous studies.
Overall, G14.33$-$0.64 shows no significant peculiar motion and is consistent with the circular Galactic rotation model.

\vspace{5mm}
We are deeply grateful to an anonymous referee for his/her careful reading of the paper and for a number of invaluable suggestions.
We would like to express our sincere gratitude to all staff members and students at VERA and Kagoshima University for their continuous support. 
MS gratefully acknowledges the financial support from the Research Fellowships of the Japan Society for the Promotion of Science (JSPS) for Young Scientists.

\end{document}